\documentclass{an}
\usepackage{graphicx}
\usepackage{times}
\Volume{01}              
\Year{2002}              
\Month{10}               
\Pagespan{223}{227}      

\def\xmm{{XMM-Newton~}}

\newcommand{\oergs}[1]{$10^{#1}$ erg s$^{-1}$}
\newcommand{\nh}{\hbox{$N_{\rm H}$}}

\begin{document}
\lhead[\thepage]{A.N. Author: Title}
\rhead[Astron. Nachr./AN~{\bf XXX} (200X) X]{\thepage}
\headnote{Astron. Nachr./AN {\bf 32X} (200X) X, XXX--XXX}

\title{Deep \xmm survey of M33\thanks{This work is based on observations 
    obtained with XMM-Newton, an ESA Science Mission 
    with instruments and contributions directly funded by ESA Member
    States and the USA (NASA).}}

\author{W. Pietsch\inst{1} 
\and  M. Ehle\inst{2}
\and  F. Haberl\inst{1}
\and  Z. Misanovic\inst{1}
\and  G. Trinchieri\inst{3}}
\institute{
Max-Planck-Institut f\"ur extraterrestrische Physik, Giessenbachstra\ss e,
D-85741 Garching, Germany 
\and 
\xmm Science Operations Centre, Apdo. 50727, E-28080 Madrid, Spain
\and 
Osservatorio Astronomico di Brera, via Brera 28, I-20121 Milano, Italy}

\date{Received {\it date will be inserted by the editor}; 
accepted {\it date will be inserted by the editor}} 

\abstract{In an \xmm raster observation of the bright local group spiral galaxy
M33 we study the population of X-ray sources (X-ray binaries, supernova
remnants, super-shells) down to a 0.5--10 keV luminosity of \oergs{35} --
more than a factor of 10 deeper than earlier ROSAT observations. EPIC spectra
and hardness ratios are used to distinguish between different source classes. We
confirmed the 3.45 d orbital light curve of the X-ray binary M33 X7, detected a
transient super-soft source in M33, and searched for short term variability of
the brighter sources. We characterize the diffuse X-ray component that is 
correlated with the inner disk and
spiral arms. We will compare the results with other nearby galaxies.
}

\correspondence{wnp@mpe.mpg.de}

\maketitle

\section{Introduction}
The Local Group Sc galaxy M33 at a distance of 795 kpc (van den Bergh
1991) with relatively low inclination of 56$^{\circ}$ (Zaritzky, Elston \& Hill 
1989) and its moderate Galactic foreground absorption 
(\nh$ = 6\,10^{21}$\,cm$^{-2}$, Stark et al. 1992) is
ideally suited to study the X-ray source population and diffuse emission in a
nearby spiral. The Einstein X-ray Observatory detected diffuse emission from hot
gas in M33 and 17 unresolved sources (Long et al. 1981; Markert \& Rallis 1983;
Trinchieri, Fabbiano \& Peres 1988). First ROSAT HRI and PSPC observations
revealed 57 sources and confirmed the detection of diffuse X-ray emission which
may trace the spiral arms within 10' radius around the nucleus (Schulman \& Bregman 1995; 
Long et al. 1996). Combining all archival ROSAT observations of the field, Haberl
\& Pietsch (2001; HP01 hereafter) found 184 X-ray sources within 50' radius
around the nucleus,
identified some of the sources by correlations with previous X-ray, optical and
radio catalogues, and in addition classified sources according to their X-ray
properties. They found candidates for super-soft X-ray sources (SSS), 
X-ray binaries (XRBs), supernova remnants (SNRs), foreground stars and active
galactic nuclei (AGN) in the background.

Two M33 sources are known for their outstanding X-ray properties (Peres et
al. 1989). The brightest source (X8, luminosity of about 
$10^{39}$\,erg~s$^{-1}$) is the most luminous X-ray source in the Local Group
of galaxies and coincides with the optical center of M33. Its time variability  
(Dubus et al. 1997), its point-like nature
seen by ROSAT HRI (Pietsch \& Haberl 2000) and Chandra (Dubus \& Rutledge 2002) 
and its X-ray spectrum best described by an absorbed power law plus disk blackbody model 
(e.g. Ehle, Pietsch \& Haberl 2001, E01 hereafter; 
La Parola et al. 2002) point at a black hole
XRB. A possible periodicity of 106 days was not
confirmed in later observations (see Parmar et al. 2001). The second source (X7)
is an eclipsing XRB  with a binary period of 3.45 d and
possible 0.31 s pulsations (Schulman et al. 1993, Dubus et al. 1997, 1999; D99
hereafter; Larson \& Schulman 1997).

Here we present \xmm raster observations which were carried out within the
telescope scientist guaranteed time program to survey M33 homogeneously for
point sources with a sensitivity of $10^{35}$\,erg~s$^{-1}$ in the 0.5--10 keV
band, a factor of ten deeper than previous surveys. First X-ray false color images combining 
different energy bands have been discussed in Pietsch (2002). 

\section{Observations, data analysis and results}
\begin{figure}
\resizebox{\hsize}{!}
{\includegraphics[bbllx=37pt,bblly=130pt,bburx=572pt,bbury=650pt,clip]{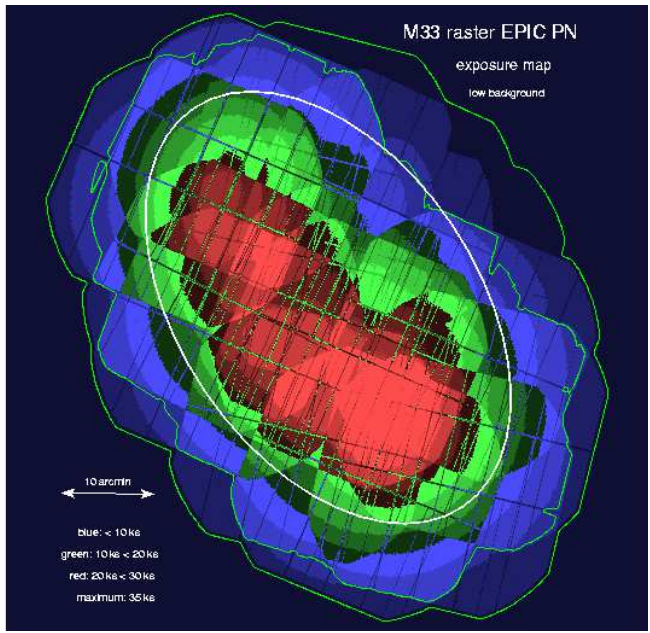}}
\caption{\xmm EPIC low-background exposure map of the M33 raster. Contours are
at 0 and 5 ks, gray-scale steps every 2 ks. The maximum exposure is 35 ks in the SW M33 
disk. The optical extent of M33 is marked by the white D$_{25}$ ellipse.}
\label{expo}
\end{figure}
Fifteen 10 ks observations on M33 were scheduled with \xmm (Jansen et al. 2001) 
between August 2000 and January 2002. The first 
two observations were 
performed with the thick filter in front of the EPIC detectors (Turner et al.
2001; Str\"uder et al. 2001) while for the rest 
the medium filter was used. 
The raster pointings with a spacing of $\sim$10'  
ensure that each position within the optical D$_{25}$ ellipse of M33 is 
covered at least 3 times. During the proposed 30 ks integration time, for each source 
inside the ellipse at least 15 counts are detected in the EPIC PN camera in the 
0.5--10 keV band (taking into account the X-ray telescope and instrument 
response and assuming a 5 keV thermal bremsstrahlung spectrum).
Several observations suffered from high 
particle background reducing in some cases the integration time useful for the 
raster imaging to below 1.3 ks. Fig.~\ref{expo} shows the 
EPIC PN coverage of M33 (total integration time 112 ks, 92 ks with medium filter). 
Several pointings have been granted for reobservation to homogenize the survey.

The data analysis was performed using tools in the SAS v5.3.3 and 
FTOOLS v5.1 software packages, the imaging applications DS9 v2.1 and KVIEW v1.1.19, the timing 
analysis package XRONOS v5.19 and spectral analysis software XSPEC v11.2.   

\subsection{X-ray images}
\begin{figure}
\resizebox{\hsize}{!}
{\includegraphics[bbllx=62pt,bblly=54pt,bburx=464pt,bbury=525pt,clip]{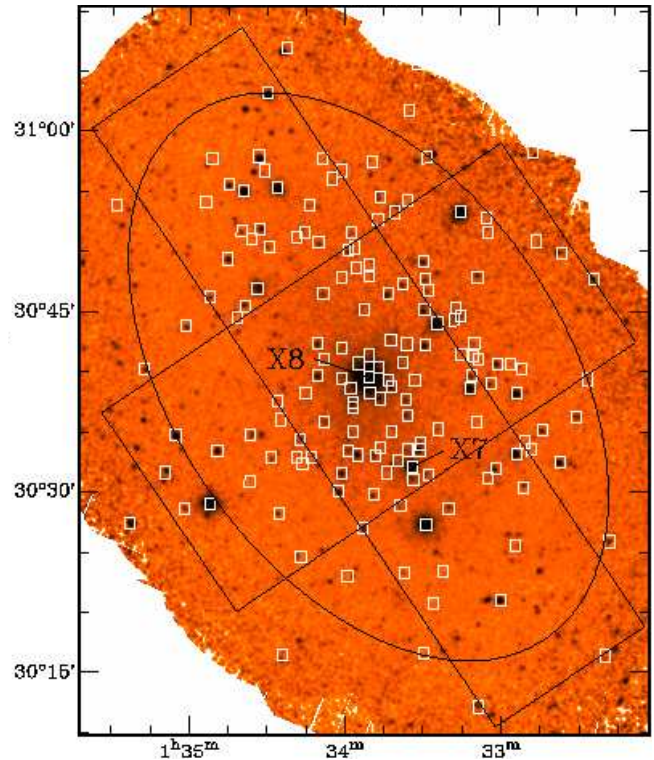}}
\caption{\xmm EPIC low-background 
image of M33 (0.2--4.5 keV). Data of the EPIC PN, MOS1 and MOS2 cameras have 
been combined (see text). White
squares indicate ROSAT sources from the HP01 catalogue. The optical extent 
of M33 is marked by the black D$_{25}$ ellipse, the extraction areas for Fig.~\ref{proj} as black boxes.}
\label{m33broad}
\end{figure}
The creation of the images of the raster survey of M33 was performed in several steps.
We first cleaned the EPIC event lists of the individual observations for bad pixels and 
times of high background and calculated sky positions with respect to a common
reference point close to the center of the galaxy. We then accumulated images and exposure maps 
for the individual observations in the energy bands 0.2--0.5 keV (B1), 0.5--1.0
keV (B2), 1.0--2.0 keV (B3), 2.0--4.5 keV (B4), and
4.5--12 keV (B5). For EPIC PN we also created "out of time event" 
(OOT) images. For EPIC PN we excluded the energy range 7.2--9.2 keV from the B5 
band images, as in this subband internal florescence lines lead to a spatially 
very inhomogeneous 
background. We smoothed all images with a Gaussian of 12.6" FWHM, subtracted OOT 
images for EPIC PN, added them up and corrected for exposure times. 

The 0.2--4.5 keV image of M33 including all EPIC instruments (Fig.~\ref{m33broad}) shows many 
more sources than previously detected with ROSAT within the D$_{25}$ ellipse and indicates 
unresolved emission surrounding the bright source (X8) close to the nucleus and
from the optically bright inner disk and southern spiral arm. 

\subsection{Diffuse emission from the inner disk}
\begin{figure}
\resizebox{\hsize}{!}
{\includegraphics[bbllx=54pt,bblly=95pt,bburx=222pt,bbury=436pt,angle=-90,clip]{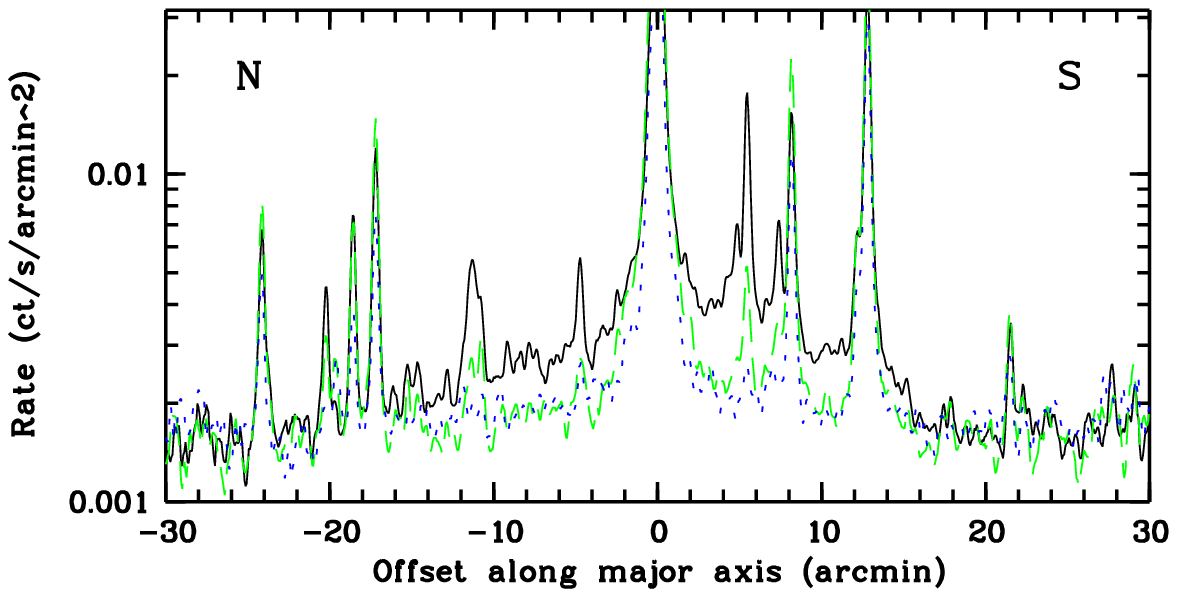}}
\resizebox{\hsize}{!}
{\includegraphics[bbllx=54pt,bblly=95pt,bburx=222pt,bbury=436pt,angle=-90,clip]{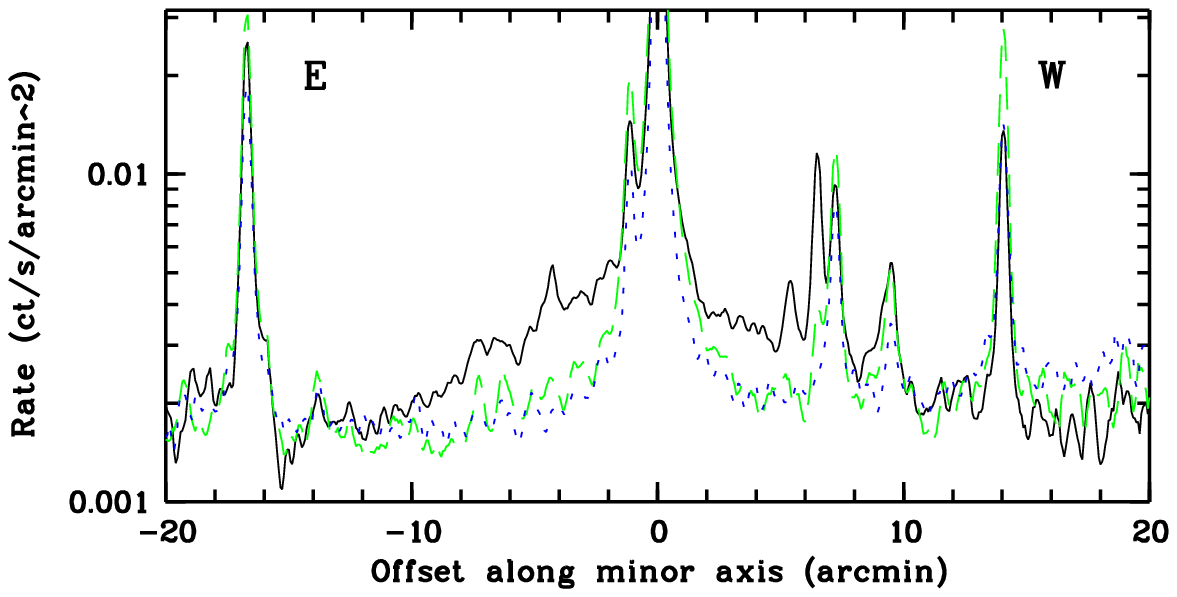}}
\caption{\xmm EPIC PN emission profile along the major (width 15', top) and minor
(width 20', bottom) axes in the (0.5-1.0; solid line), (1.0--2.0; dotted line),
(2.0--4.5;dashed line) keV band.}
\label{proj}
\end{figure}

To further investigate the diffuse emission we extracted emission profiles in
rectangular boxes along the major and the minor axes of M33 (width of 15' and 20', respectively) 
from individual energy band images of the EPIC PN (Fig.~\ref{proj}).
The bright point sources can clearly be identified as peaks in the plots. 
While in the B3 and B4 band the background level beyond the point sources is 
rather flat, there is an enhancement of emission in the B2 band both within $\sim$15' 
from the galaxy center along the major axis and $\sim$10' from the galaxy 
center along the minor axis profiles. This excess seems stronger to the south 
and to the east of the nucleus. While 
this in principle may be caused by unresolved point sources that are predominantly
radiating in the soft band (like SSS, SNRs) the more plausible interpretation is 
emission from hot gas in the interstellar medium (ISM) of the M33 disk or halo. Such
a component would be similar to the hot ISM measured in the LMC and SMC (see
Sasaki, Haberl \& Pietsch 2002). A detailed
spatial and spectral analysis, cutting out point sources, is in progress. It is however 
clear that this will be no easy task due to the extended point spread function (PSF) of
the \xmm telescopes at large off-axis positions which heavily smear out the bright 
point source near the nucleus if observed at larger off-axis angles. The
investigation of the diffuse emission will therefore have to concentrate on 
individual pointings where the dominant source X8 is favorably positioned (ie.
close to the on-axis position). 

\subsection{Catalogue of point sources}
In a first attempt we produced a catalogue of X-ray point sources 
in the M33 field with the source detection programs 
of the SAS. We created unsmoothed merged images, exposure maps and masks in a similar way 
as described in Sect. 2.1. Background maps were calculated taking into account
OOT events for EPIC PN. 
In the overlapping fields the PSF at a given sky position is made up by the
overlay of the individual PSFs according to the different off-axis 
parameters. The effective PSF for a source in the merged image therefore 
certainly differs from the one that the SAS calibration would assume for the 
offset angle from the assumed reference point of the raster. 
To circumvent this effect we used a calibration file with
a PSF similar to a source at an off-axis angle of 6' for the total field. This is a first
approximation for the PSF of source counts collected at a wide range of off-axis angles, and
should give reasonable count rate estimates. More sophisticated
approaches are certainly needed. 

Simultaneous detection runs included images of all energy bands of the individual EPIC 
instruments. We also combined all EPIC instruments, using simultaneously 
5$\times$3 images 
(five energy bands for EPIC PN, MOS1 and MOS2). This
resulted in a catalogue of source positions, with count rates and hardness 
ratios. 

We detected more than 400 sources in the overlapping field of view of the EPIC 
instruments. While for the fainter sources one can determine 
position, flux and possibly hardness ratios (HRs), 
the brighter sources ($>500$ counts)
allow in addition the investigation of their spectra and time variability. 
Detailed discussion of the source catalogue and individual sources goes beyond 
the scope of this paper. In the following we give some first highlights.

\subsection{Spectra and time variability of bright sources}
\begin{figure}
\resizebox{\hsize}{!}
{\includegraphics[bbllx=53pt,bblly=110pt,bburx=224pt,bbury=434pt,angle=-90,clip]{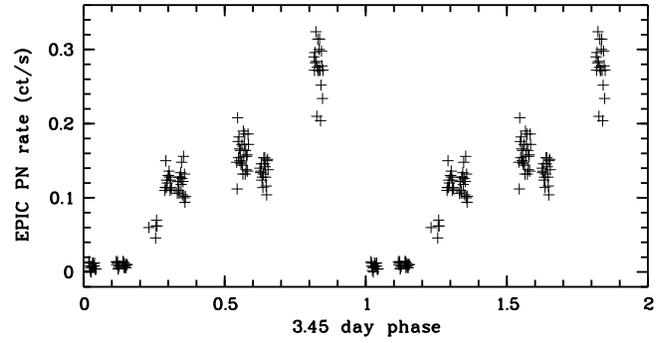}}
\caption{\xmm EPIC PN 0.3--7.2 keV light curve of the XRB M33 X7 
(integration time 500 s)
folded over the 3.45 d orbital period of D99.}
\label{rate_x7}
\end{figure}
Based on the EPIC PN data, we performed spectral fits of the brightest sources
and produced light curves to start the source identification. In E01 
we gave first results on the black hole XRB M33 X8 including an analysis of RGS spectra
of this source. Also shown are EPIC PN spectra of the XRB M33 X7, 
a SNR, and a transient short time variable SSS candidate that was bright in
August 2000 and no longer detectable in July 2001 and January 2002, a behavior
often seen in SSS (Kahabka \& van den Heuvel 1997). Many
sources that were not discussed in E01, vary in intensity between observations.   

The XRB M33 X7 was covered by eight EPIC PN fields. 
EPIC PN counts of the source region during 
low background times were integrated over 500 s and folded over the 3.45 d orbital
period (D99). In Fig.~\ref{rate_x7} variability on 500 s
time scale as well as on the orbital period is clearly visible. The source
intensity seems to increase towards the end of the orbit. This can either
reflect real orbital or long term variability of X7 as the data in the 
individual phase blocks are separated by many orbital cycles.
The time of eclipse seems to
be shifted to later phases by 0.1, pointing at a slightly longer orbital period
than given by D99.
Better phase coverage of the X-ray eclipse will allow us to significantly improve
on the orbital parameters. In a pulsation search we could
not confirm the 0.31 s pulsation period and did not find any other periods.

\section{Source identification and classification}
\begin{figure}
\resizebox{6.6cm}{!}
{\includegraphics[bbllx=50pt,bblly=90pt,bburx=368pt,bbury=433pt,angle=-90,clip]{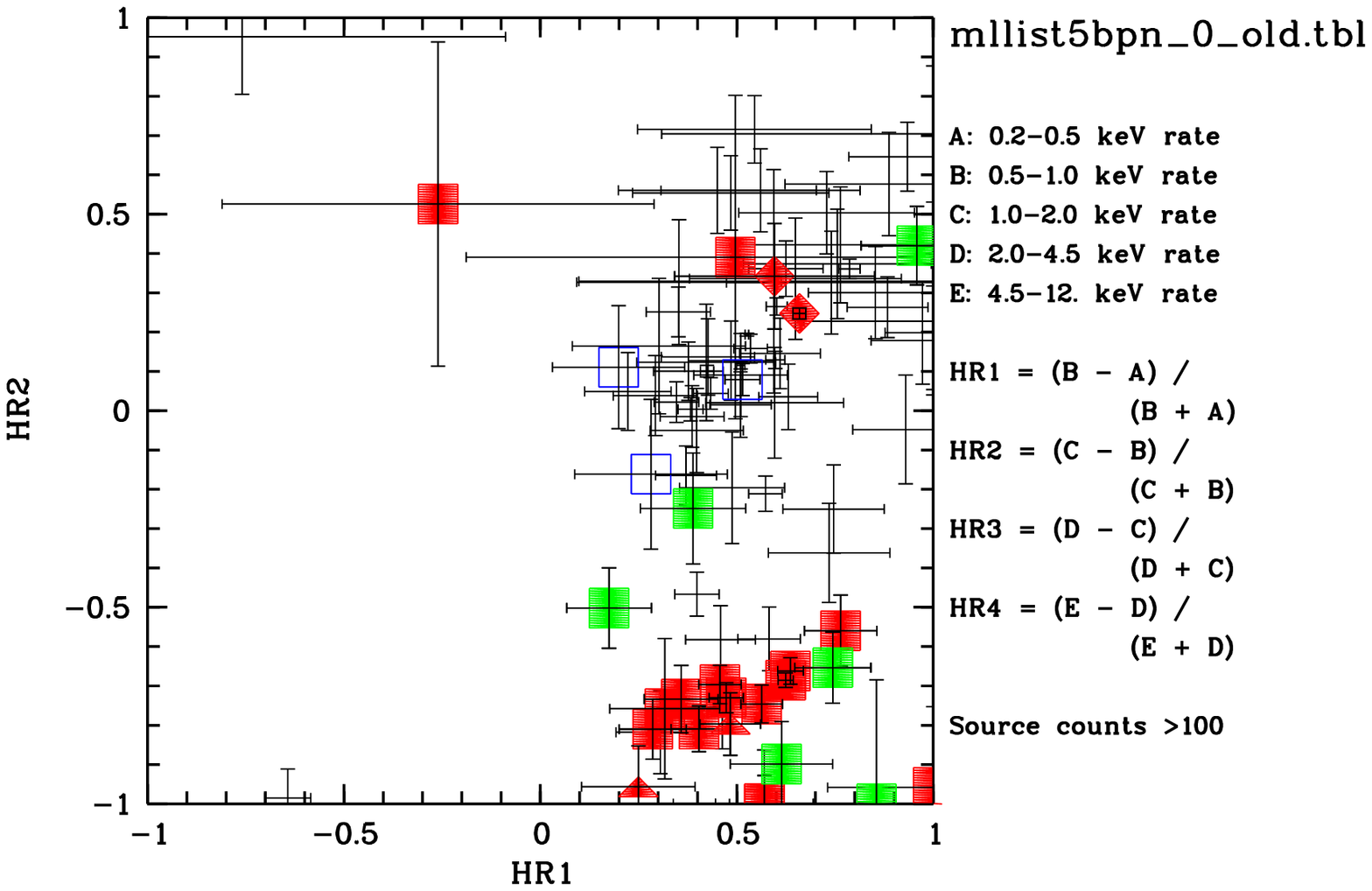}}
\resizebox{6.6cm}{!}
{\includegraphics[bbllx=50pt,bblly=90pt,bburx=368pt,bbury=433pt,angle=-90,clip]{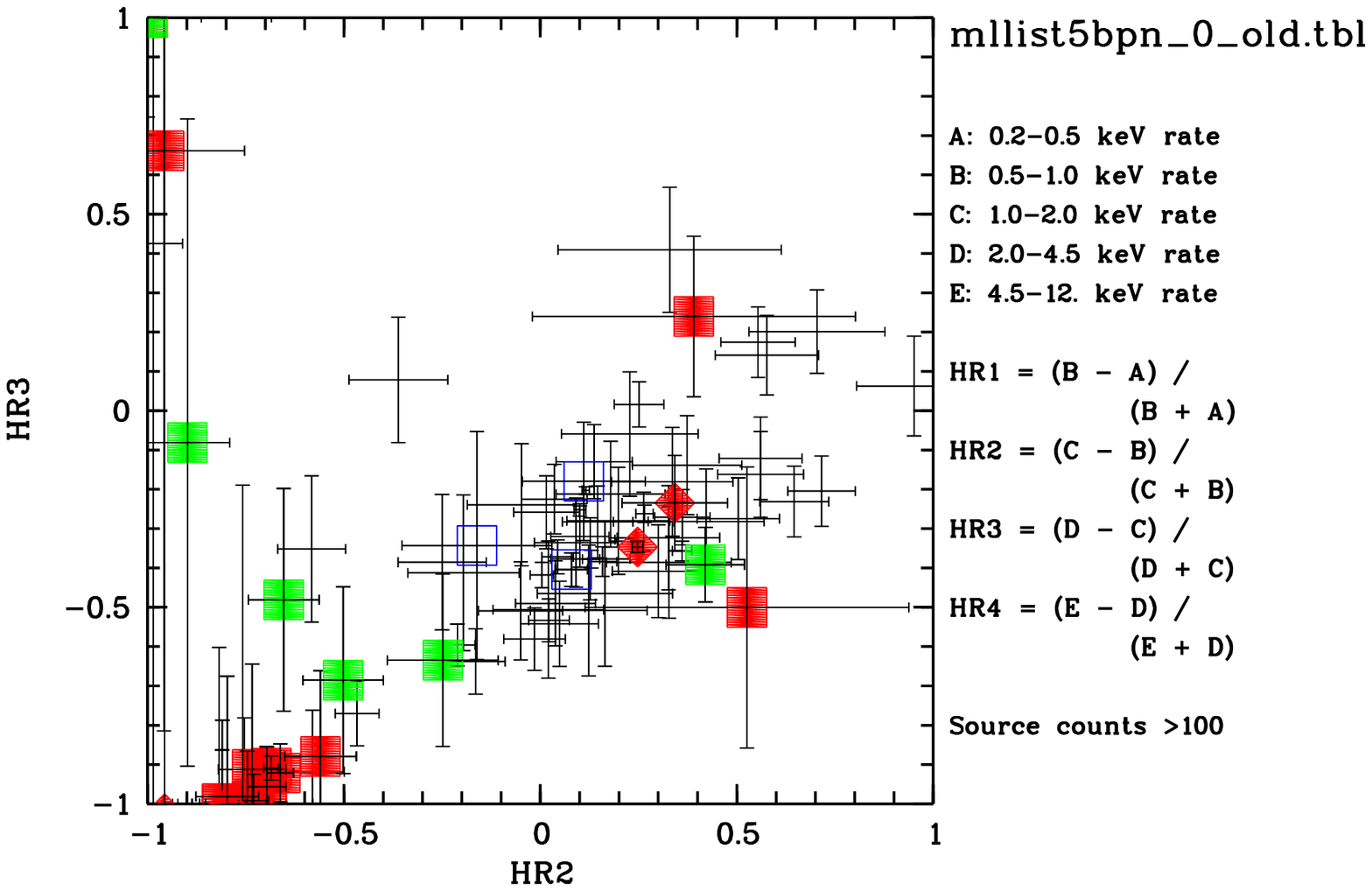}}
\caption{\xmm EPIC PN hardness ratio plots: (top) HR2 against HR1, (bottom)
HR3 against HR2. Shown are only the 75 sources with more than 100 counts and/or 
ROSAT classifications. Two candidates for XRBs are marked as filled dark lozenges,
2 for SSSs as filled triangles, 15 for SNRs as filled dark squares, 6 for foreground stars as 
filled bright squares, 3 for AGN as open squares.}
\label{hr}
\end{figure}
In our ROSAT work on Local Group galaxies we developed a scheme to classify
detected
sources by using their X-ray properties (mainly hardness ratios and extent, see 
HP01). With \xmm we are broadening this method in adding information from the
harder energy bands accessible with EPIC. We define hardness ratios as
HR1=(b2-b1)/(b2+b1), HR2=(b3-b2)/(b3+b2), HR3=(b4-b3)/(b4+b3),
HR4=(b5-b4)/(b5+b4), where "bi" is the
count rate in band "Bi". Specifically the use of HR3 and HR4 
should allow us to separate XRBs
from AGN, in addition to the classification already possible with ROSAT. With the
higher sensitivity of \xmm  we also will be able to classify fainter 
sources. As a first attempt, Fig.~\ref{hr} shows X-ray "color/color" hardness
ratio  
plots for the 75 sources in the EPIC PN catalogue with more than 100 counts 
and/or with ROSAT classification. ROSAT SNRs, SSS, XRBs, foreground 
star and AGN candidates are in general clearly separated. However, there seem to be
several miss-classifications within the ROSAT list or miss-identifications as
some of the sources do not fall into the regions in the 
EPIC PN hardness ratio diagrams populated by
these source classes. Cross-correlations with the catalogue of
Gordon et al. (1999) yielded several new X-ray SNR and AGN candidates in the
field. More detailed investigations are in progress.

\section{Conclusions}
As demonstrated above, our setup of \xmm raster observations of M33 yields very
interesting results, fully confirming our expectations. We have detected more 
than 400 sources down to a luminosity limit which is below \oergs{35} if the
sources are located in M33. We 
showed the possibilities to classify them from their X-ray properties alone. There 
is unresolved emission detected at energies below 1 keV, most likely from hot gas in the inner 
region of the disk and possibly from the southern spiral arm and/or halo above, 
that needs to be followed up.  

Several of the X-ray observations were heavily affected by high background and 
will be re-scheduled to homogenize the survey. Identifications of X-ray sources 
with surveys and follow-up observations at other wavelengths are in progress. 

\acknowledgements
The \xmm project is supported by the Bundesministerium f\"{u}r
Bildung und Forschung / Deutsches Zentrum f\"{u}r Luft- und Raumfahrt 
(BMBF/DLR), the Max-Planck Gesellschaft and the Heidenhain-Stiftung.

\end{document}